\documentclass[12pt]{article}

\usepackage{graphicx}

\begin{document}

\title{UNBOUND GEODESICS FROM THE ERGOSPHERE AND THE M87 JET PROFILE}
\author{J. Gariel$^2$\thanks{e-mail: jerome.gariel@upmc.fr}, G. Marcilhacy$^2$\thanks{e-mail: gmarcilhacy@hotmail.com}
{and N. O. Santos$^{2,3}$\thanks{e-mail: nilton.santos@upmc.fr}}\\
{\small {$^2$LERMA-UPMC, Universit\'e Pierre et Marie Curie, Observatoire de
Paris,}}\\
{\small {CNRS, UMR 8112, 3 rue Galil\'ee, Ivry sur Seine 94200, France}}\\
{\small {$^3$School of Mathematical Sciences, Queen Mary, University of
London,}}\\
{\small {Mile End Road, London E1 4NS, UK}}}
\maketitle

\begin{abstract}
Assuming that the spin $a$ of the black hole, presumably located at the core of the active galactic nuclei Messier 87 (M87), takes the value which maximises the ergospheric volume of the Kerr spacetime, we find results compatible with the recent observations obtained by high resolution interferometry on the origin of the jet, which would be located inside the innermost stable circular orbit diameter. Moreover, we find a flow of unbound geodesics issued from the ergoregion able to frame the best fits at large scale recently obtained for describing the observed profile of the relativistic jet launched from this central engine.
\end{abstract}

\section{The jet model}

The arrival of high resolution observations, of the order of microseconds of
arc, will allow us soon to observe the origin of extragalactic jets and
specially the less distant ones like the M87. These data should tip the balance in favor of one of the two main kinds os models. The first ones situate the origin of the jet close to the
interior of the accretion disk, near the innermost stable
circular orbit (ISCO) (Blandford \& Znajek 1977). While the second ones consider that the
jet originates inside the ergosphere of the Kerr black hole (BH) situated at the
centre of the galaxy, for instance proposed in (Williams 1995, 2004) and
(Gariel et al. 2010; Pacheco et al. 2012).

The model (Gariel et al. 2010; Pacheco et al. 2012) is based on the unbound particle geodesics
leaving the Kerr BH ergosphere. These particles are produced by a Penrose
process (Penrose 1969) inside the ergosphere and, while ejected, follow
asymptotic geodesics parallel to the axis of symmetry $z$ at
a radial distance
\begin{equation}
\rho =\rho _{1}=a\left[1+\frac{\mathcal Q}{a^2(E^2-1)}\right]^{1/2}, \label{1a}
\end{equation}
where ${\mathcal Q}$ is the Carter constant, $a$ the BH spin and $E$ the particle energy.
There is a limited number of discrete
values $\rho_1$ for which the energy $E$ of the particles tends to infinity,
while near the neighborhood of each of these values the energy diminishes
steeply. This permits modeling a thin jet highly energetic and collimated.

\section{Maximal ergospheric volume}

Admitting that the Penrose process is the source of production of particles
building the jet, it is reasonable to suppose that its frequency increases
proportionally to the ergospheric volume, which is the only region where it
can take place. Hence, the larger is the ergospheric volume the larger is
the amount of particles produced to take part of the jet, and the more
powerful becomes the jet.

The trace of the ergosphere in the plane $(\rho,z)$ is a closed curve, with equation
\begin{equation}
z^{2}=\left[ 1-a^{2}\left( 1-\frac{\rho }{a}\right) \right] \left( 1-\frac{\rho }{a}\right) ,  \label{1}
\end{equation}
depending only on the parameter $a$, the Kerr BH angular momentum per unit mass and where
we normalized its mass $M=1$. The surface $S$ inside this curve can be
calculated from
\begin{equation}
S=\int_{0}^{a}\left\{ \left[ 1-a^{2}\left( 1-\frac{\rho }{a}\right) \right]
\left( 1-\frac{\rho }{a}\right) \right\} ^{1/2}d\rho ,  \label{2}
\end{equation}
which, after integration, produces
\begin{equation}
S=\frac{1}{8a^{2}}\left[ \frac{\pi }{2}-\arcsin (1-2a^{2})\right] -\frac{1}{%
4a}(1-2a^{2})(1-a^{2})^{1/2}.  \label{3}
\end{equation}
We can plot the function $S(a)$, figure 1, for $a\in\lbrack 0,1]$ and
obtain the value of $a$ where $S$ is maximal:
\begin{equation}
a=0.877004.  \label{4}
\end{equation}
The corresponding trace of the ergosphere, (\ref{1}), for $a$ given by (\ref{4}), generating the maximal ergospheric volume, is plotted in figure 2.

\begin{figure}[ht]
\centering
\centerline{\includegraphics[width=10cm]{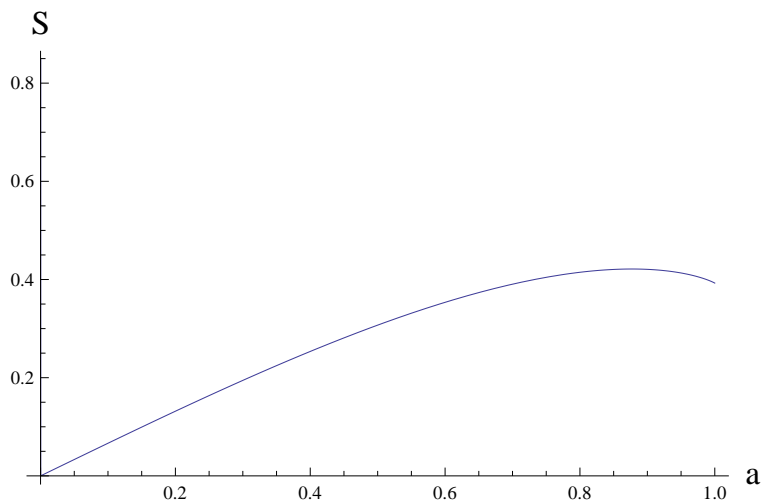}}
\caption{\small{Plot of the ergoregion projection $S$ in the $(\rho,z)$ plane as a function of BH spin $a$ (see (\ref{3})). Its maximum $S_{max}=0.421385$ is for $a=0.877004$.}}
\label{figure1}
\end{figure}
\begin{figure}[ht]
\centering
\centerline{\includegraphics[width=10cm]{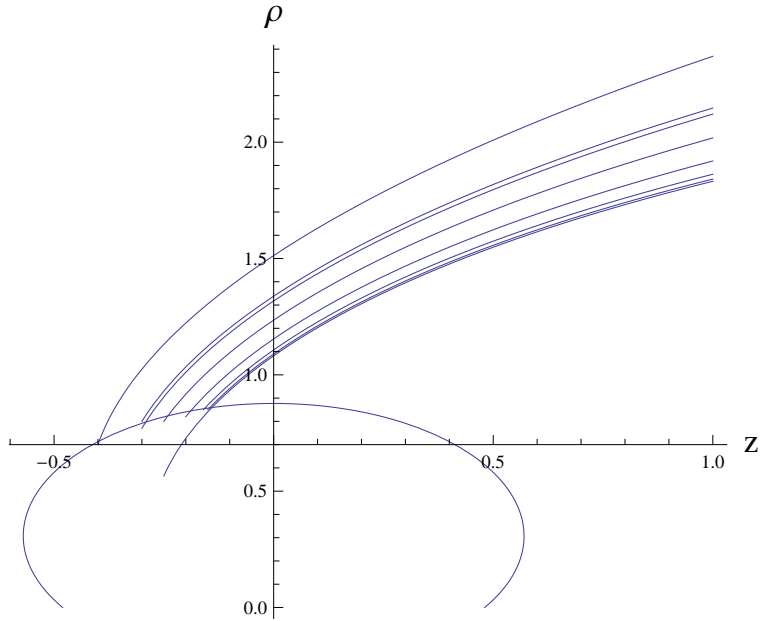}}
\caption{\small{Plot of the ergosphere trace, in the $(\rho,z)$ plane, limiting the maximal ergoregion obtained in figure 1 for $a=0.877004$. We also plot the beginning of some unbound geodesics with high energy, $E=10^6$, from this maximal ergosphere, of interest for modelling the M87 jet, as we can see in sections 4 and 5. The geodesics asymptotic to $\rho_1=3.1637$ is obtained for the i.c. $(\rho_i=0.565256,z_i=-0.25)$ (see section 4), and a set of imperfectly collimated near geodesics framed by two geodesics, and "external" one with i.c. $(\rho_i=0.7,z_i=-0.4)$, and an "internal" one with i.c. $(\rho_i=0.85,z_i=-0.15)$ (see section 5).}}
\label{figure2}
\end{figure}

\section{Jet produced by M87}

The first observation of an extragalactic jet dates back to the year 1918
and concerned the one produced by M87, being very long and
powerful. Currently, the M87 nucleus distance is estimated at $16.7\pm 0.6$ Mpc (Blakeslee et al. 2009) and the mass of its central BH to be $M=(6.2\pm 0.4)\times 10^9M_{sun}$ (Gebhardt et al. 2011). Its relativistic jet extends for hundreds of kpcs before ending in large radiolobes.

In the framework of the model discussed in (Gariel et al. 2010; Pacheco et al. 2012), it is
natural, according to the results of the section 2, to expect that the most
energetic jets produced by a Kerr BH of mass $M$ are obtained for $a$ given
by (\ref{4}). It is natural to assume too that the jet produced by M87 fits
into this model. We call the attention that this value for $a$ is consistent
with the few evaluations of the M87 BH spin obtained to date, like $a>0.65$ in
(Wang et al. 2008), and more recently $a>0.8$ in (Li et al. 2009).

Hence we look for particles moving along unbounded geodesics with high
energy, we take $E=10^6$, produced in the ergoregion, which means, in 2D,
produced between the ergosphere and the event horizon situated on the $z$
axis. These particles move asymptotically parallel to the $z$ axis with
$\rho\rightarrow\rho_1$ when $z\rightarrow\infty$. In fact, there are other
geodesics, near these perfectly collimated ones, that asymptotically behave
$\rho\rightarrow\infty$ when $z\rightarrow\infty$, however they are not very
divergent, i.e., they satisfy $\rho\ll z$ for a long distance along
the axis.

Now we need to determine the parameter $\rho_1$. In order to do this we
consider the recently obtained high resolution data (Doeleman et al. 2012).

\section{Jet near the core of M87}

The data presented in (Doeleman et al. 2012) opens a new era of resolutions never
before achieved, which are of the order tens of microseconds of arc.
Hopefully soon it can be expected resolutions even better, of microseconds
of arc. With these observations the models considering jets produced by the
accretion disk near the ISCO (Blandford \& Znajek 1977) and models with jets leaving
the ergosphere (Williams 1995, 2004; Gariel et al. 2010; Pacheco et al. 2012) will be
discriminated. Among these models the one that is based only on gravity to explain the genesis of jets considers that the Kerr spacetime has the essential features to explain their formation, acceleration and collimation (Gariel et al. 2010; Pacheco et al. 2012). Other fields, excepting the
gravitational, like the magnetic field are no more that a complement for
their formation.

Through interferometry by using the VLBI array and by choosing a frequency
of the order 229 Ghz (or wave length 1.3 mm) allowing a good transparency up
to the core\footnote{We call the "core" the region where the jet is produced, which means
extending maximally up to the radius of the ISCO and, consequently,
including the ergosphere. This "core" has an altitude $z$ of the order of $M$.
We precise this point since it is not always understood in this way. For
instance in (Hada et al. 2011) it is defined centred transverse to the axis $z$.
Hence there are "cores" for each altitude $z$ according to its observed
frequency} of the galaxy and with unprecedent resolution of the order of
$40\pm1.8 \mu$ as corresponding\textbf{, }for a distance 16.7$\pm$0.6 Mpc, to
a length of the order of 5.5 $r_{S}$, Doeleman et al. (Doeleman et al. 2012)
evaluated the maximal diameter of the emerging jet at
\begin{equation}
r_{jetApp}=5.5\pm 0.4r_{S},  \label{5}
\end{equation}
where $r_{S}=2M$ is the Schwarzschild BH radius.

Then, by admitting the validity of the model by Blandford and Znajek (Blandford \& Znajek 1977),
they assume that the jet starts at the ISCO of the accretion
disk and put
\begin{equation}
r_{jetApp}=r_{ISCOApp}.  \label{6}
\end{equation}
This assumption implies too that the origin of this observation is on the
plane of ISCO, i.e. the equatorial plane $z=0$. The apparent radius of ISCO,
plotted in (Doeleman et al. 2012) as a function of the spin $a$ (see figure 3 in
(Doeleman et al. 2012)), is the intrinsic radius corrected by the effect of
gravitational lensing caused by its proximity to the core endowed with a
Kerr BH. In this figure, the horizontal line representing $r_{jetApp}$
intersects the prograde radius of ISCO, which allows to evaluate the spin of
the BH as $a\approx 0.625\pm 0.25$ and to show that only the prograde part
is viable. We observe that the so obtained value of $a$ is inside the limit of
(Wang et al. 2008), but outside the limit of (Li et al. 2009).

Now we interpret the same result in the light of the model given by (Gariel et al. 2010; Pacheco et al. 2012).
Instead of considering the "apparent" distances, we deduct
the lensing and consider the "intrinsic" distances
\begin{equation}
r_{jetInt}\approx r_{jetApp}-1.3,  \label{7}
\end{equation}
by observing that the curve $r_{ISCOInt}(a)$, figure 3, roughly is translated downwardly
about 1.3 from the $r_{ISCOApp}(a)$ given in
figure 3 in (Doeleman et al. 2012).
\begin{figure}[ht]
\centering
\centerline{\includegraphics[width=10cm]{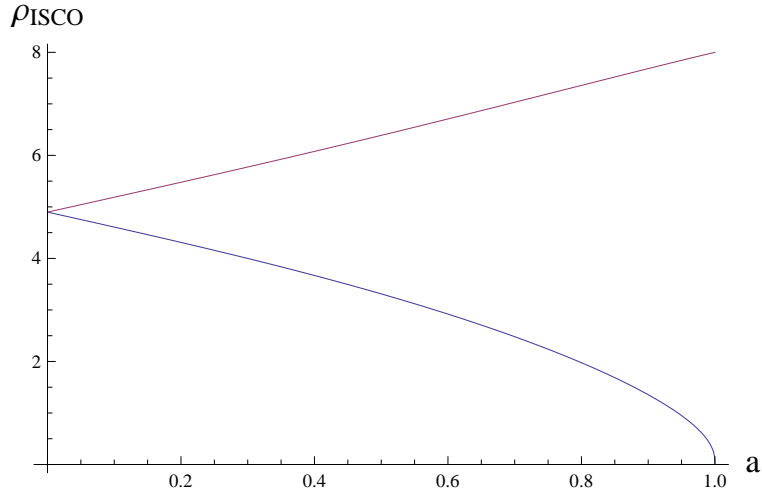}}
\caption{\small{Plot of the Kerr prograde and retrograde ISCO for $\rho_{ISCO}$ as function of the BH spin $a$ obtained from the transformation formula (\ref{8}) in the equatorial plane $\theta=\pi/2$, with the standard $r_{ISCO}=3+Z_2\pm[(3-Z_1)(3+Z_1+2Z_2)]^{1/2}$, with $Z_1=1+(1-a^2)^{1/3}[(1+a^2)^{1/3}+(1-a^2)^{1/3}]$ and $Z_2=(3a^2+Z_1^2)^{1/2}$.}}
\label{figure3}
\end{figure}
Writing these quantities in terms of Weyl
coordinates, given by (1) in (Pacheco et al. 2012), we obtain
\begin{equation}
\rho
_{jetInt}=[(r_{jetInt}-1)^{2}-(1-a^{2})]^{1/2}=[(r_{jetApp}-2.3)^{2}-(1-a^{2})]^{1/2}.
\label{8}
\end{equation}

Now we can plot the curves $\rho _{jetInt}(a)$ for values in the range
$r_{jetApp}\in \lbrack 5.1,5.9]$, which are given in figure 4.
\begin{figure}[ht]
\centering
\centerline{\includegraphics[width=10cm]{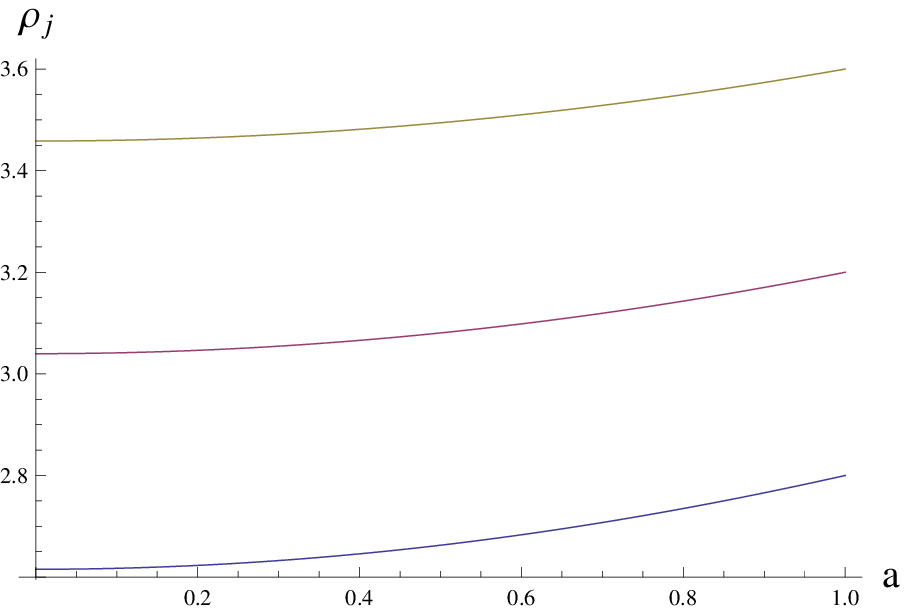}}
\caption{\small{The central curve is the plot of the intrinsic (transverse) radius $\rho_{jetInt}$ (here noted $\rho_j$) as function of $a$, deduced from the apparent (radial) radius $r_{jetApp}=5.5$ observed at the jet base by (Hada et al. 2011). The two intrinsic transverse radii corresponding to the two limits of their margin of error, $r_{jetApp}\in[5.1,5.9]$, are also plotted. For the value $a=0.877004$, corresponding to the largest ergoregion (see figure 1), the radius at the base of the jet is $\rho_j=3.16372$.}}
\label{figure4}
\end{figure}
In particular,
from the central curve, corresponding to the apparent radius $r_{jetApp}=5.5$,
we deduce for the maximal volume of the ergoregion (\ref{4})
\begin{equation}
\rho _{jetInt}=3.1637.  \label{9}
\end{equation}
We abandon the hypothesis (\ref{6}) assumed by \cite{Doeleman} and we
replace it by
\begin{equation}
\rho _{jetInt}=\rho _{1}.  \label{10}
\end{equation}
This allows us to obtain the unbound $\rho_1$-asymptotic geodesic, (see the
lower curve, asymptotically parallel to the $z$-axis, in figure 5), with
parameters $a=0.87705$, $E=10^{6}$ and $\rho_1=3.1637$ after numerical
integration of (21) given by (Gariel et al. 2010) and initial conditions (i.c.)
\begin{equation}
\rho_i=0.56524, \;\; z_i=-0.25,  \label{11}
\end{equation}
which are well inside the ergosphere (see figure 2).

\section{Jet far from the core of M87}

Observations concerning the geometry of the jet far from the core of the
galaxy M87 have been performed first by Junor et al. (Junor et al. 1999) and being
interpreted, like with the observations of Doeleman et al. (Doeleman et al. 2012),
by assuming the validity of the model of Blandford and Znajek (Blandford \& Znajek 1977). 
More recent observations, Asada and Nakamura (Asada \& Nakamura 2012),
confirm and specify the data obtained by Junor et al Junor et al. 1999).

The fact that we can interpret the observations obtained by Doeleman et al.
(Doeleman et al. 2012) in the light of model (Pacheco et al. 2012), induces us to ask if
it can say something about its geometrical structure far from the core. To
answer this question we confront the fittings suggested by Asada \& Nakamura
(Asada \& Nakamura 2012) to the geodesics studied in the model (Pacheco et al. 2012).

Asada \& Nakamura (Asada \& Nakamura 2012)  observe an important changing of the jet slope
after a certain altitude of the order $z\approx 10^{5}$, near of the
Bondi radius, indicating a divergent tendency. From these
observations they established the following best fitting laws. For $z\in
\lbrack 10^{2},5\times 10^{5}]$,
\begin{equation}
z=0.2\left( \frac{\rho }{0.8}\right) ^{b_{1}}\;\mbox{or}\;\rho =0.8\left(
\frac{z}{0.2}\right)^{1/b_{1}};  \label{12}
\end{equation}
and for $z\in \lbrack 10^{5},2\times 10^{7}]$,
\begin{equation}
z=30\left( \frac{\rho }{0.2}\right)^{b_{2}}\;\mbox{or}\;\rho =0.2\left(
\frac{z}{30}\right) ^{1/b_{2}};  \label{13}
\end{equation}
where $b_{1}=1.73$ is called the parabolic type, and $b_{2}=0.96$ the
conical type. We plot these curves in figure 5 (see the two straight lines
framed by the two geodesics). We seek to confront them with geodesics of the
model (Gariel et al. 2012; Pacheco et al. 2012).

To keep continuity with the previous situation, near the core, studied in
section 4, we keep the parameters values $E=10^6$ and $\rho_1=3.1637$ to
look for unbound geodesics, not asymptotic to $\rho_1$, by numerical
integration of equation (21) given in (Gariel et al. 2010), for different i.c.
inside the ergosphere. We plot in fig. 5 an "external" and an "internal"
unbound geodesics, starting from the ergosphere, obtained for the i.c. \{$%
\rho_i=0.7$, $z_i=-0.4$\} and \{$\rho_i=0.85$, $z_i=-0.15$\} respectively,
which frame the two precedent best fits. The initial parts of these
geodesics can be also viewed in figure 2, where they frame a set of various
geodesics (not represented in the figure 5, for the sake of clarity) obtained
for intermediate initial conditions, all inside the ergosphere. The
"internal" geodesic starts above, very close to, the geodesic asymptotic to $%
\rho_1$ defined in (\ref{11}), and later diverges. The flow of geodesics
located between the external one and the internal one diverge also, forming
an outer shell, distinct of the perfectly collimated geodesic (\ref{11}).
There is evidence that the jets produced by blazars, are constituted of an
external envelope, called sheath, and of an interior thin beam
(Giroletti et al. 2004; De Villiers et al. 2005; Xie et al. 2012). These features appear naturally in our purely
gravitational model.

\begin{figure}[ht]
\centering
\centerline{\includegraphics[width=10cm]{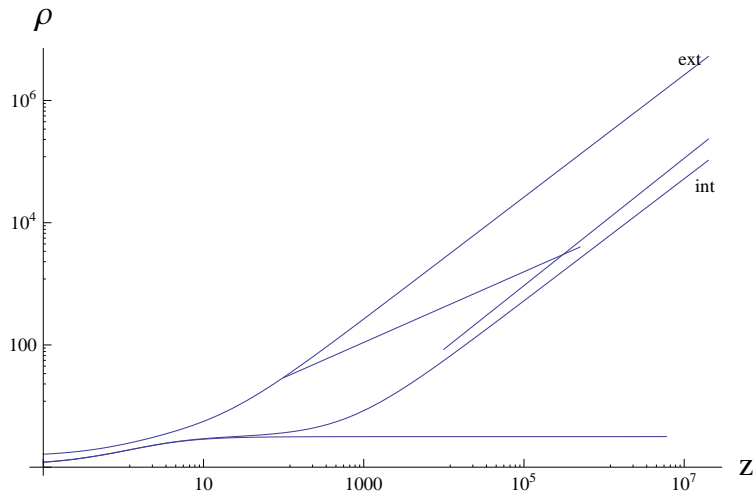}}
\caption{\small{Plots, in LogLog scales, of two diverging unbound geodesics: an "external" geodesic, obtained
for the i.c. $(\rho_i=0.7,z_i=-0.4)$,
and an "internal" geodesic with i.c. $(\rho_i=0.85,z_i=-0.15)$.
At their beginnings (see also figure 2), they are close to above the
perfectly collimated one obtained for the i.c. $(\rho_i=0.565256,z_i=-0.25)$.
The two straight lines are the plots of the
fits 1 and 2 obtained by (Asada \& Nakamura 2012) in their respective domains of
observations. We can see that the external and internal geodesics frame the
fits. Between these two geodesics, a continuous flow of geodesics can exist,
some of them being represented at their beginnings from the ergosphere in figure 2.
All of them are external to the geodesic closest to the $z$-axis
with asymptote $\rho_1=3.1637$ ("spine"), and diverge from it, shaping
the external shell of the jet ("sheath").}}
\label{figure5}
\end{figure}
We can see on figure 5 that the "conical" fit (13) is in good agreement with
the "internal" geodesic (and those close to it), the agreement seeming even
extend to lower altitudes, up to $z\sim 10^{3}$ where the fit is however not
consistent with the observations. In contrast, the "parabolic" fit (\ref{12})
does not seem able to be equated with any single geodesic of the flow. The
flow intersects the fit. That should be interpreted by noting that the
successive points emitting the observed radiations belong to the outer
portion of the jet. Thus, there is an erosion of the width of the jet during
its upward motion, mainly due to its friction with the ISM. This friction is
greater outside, where the density of the ISM is greater, because the
centrifugal force due to the BH rotation drags it outside. Without this
erosion effect, the external geodesic would remain the external part of the
jet which would remain parallel to the conical fit from $z\sim 10^3$ (or
even before), as it can be seen on figure 5.

\bigskip

\section*{References}

Asada, K., \& Nakamura, M. 2012, ApJ, 745, L28

Blakeslee, J., et al. 2009, ApJ, 694, 556

Blandford, R. D., \& Znajek, R. L. 1977, MNRAS, 179, 433

De Villiers, J. P., Hawley, J. F., Krolik, H. H., \&
Hirose, S. 2005, ApJ, 620, 878

Doeleman, S., et al. 2012, Science, 338, 355

Gariel, J., MacCallum, M. A. H., Marcilhacy, G., \& Santos,
N. O. 2010, A\&A, 515, A15

Gebhardt, K., et al. 2011, ApJ, 729, 119

Giroletti, M., Giovanni, G., Feretti, L., et al. 2004,
ApJ, 600, 127

Hada, K., et al. 2011, Nature, 477, 185

Junor, W., Biretta, J. A., \& Livio, M. 1999, Nature, 401, 891

Li, Y. R., Yuan, Y. F., Wang, J. M., Wang, J. C., \& Zhang, S. 2009, ApJ, 699, 513

Pacheco, J. A. de F., Gariel, J., Marcilhacy, G., \&
Santos, N. O. 2012, ApJ, 759, 125

Penrose, R. 1969, Rev. Nuovo Cimento (Numero Speziale), 1, 52

Williams, R. K. 1995, Phys. Rev. D, 51, 5387

Williams, R. K. 2004, ApJ, 611, 952

Wang, J. M., Li, Y. R., Wang, J. C., \& Zhang, S. 2008, ApJ, 676, L109

Xie, W., Lei, W.-H., Zou, Y.-C., et al. 2012, Res. Astron. Astrophys., 12, 817

\end{document}